\documentclass{amsart}
\usepackage{amsmath,latexsym}
\usepackage[psamsfonts]{amssymb}
\usepackage{times}
\usepackage[mathcal]{euscript}
\numberwithin{equation}{section}

\newcommand{\bbT}{\mathbb T}
\newcommand{\bbZ}{\mathbb Z}

\renewcommand{\epsilon}{\varepsilon}

\newcommand{\be}{\begin{equation}}
\newcommand{\ee}{\end{equation}}
\newcommand{\no}{\nonumber}

\newcommand{\spec}{\mathrm{spec}}
\newcommand{\C}{\mathbb{C}}

\newcommand{\F}{\mathbb{F}}

\newcommand{\N}{\mathbb{N}}

\newcommand{\R}{\mathbb{R}}

\newcommand{\T}{\mathbb{T}}

\newcommand{\Z}{\mathbb{Z}}

\newcommand{\cB}{{\mathcal B}}

\newcommand{\cE}{{\mathcal E}}
\newcommand{\cF}{{\mathcal F}}

\newcommand{\cO}{{\mathcal O}}

\DeclareMathOperator{\Ker}{\mathrm{Ker}}

{\bf}{\it}
\newtheorem{theorem}{Theorem}[section]
\newtheorem{lemma}[theorem]{Lemma}
\newtheorem{corollary}[theorem]{Corollary}
\newtheorem{hypothesis}[theorem]{Hypothesis}
\newtheorem{definition}[theorem]{Definition}
\newtheorem{proposition}[theorem]{Proposition}
\newtheorem{remark}[theorem]{Remark}
\newtheorem{example}[theorem]{Example}

\begin{document}

\title{The Threshold effects for  the two-particle Hamiltonians on
 lattices}

\author{S.  Albeverio$^{1,2,3}$}
\address{$^1$ Institut f\"{u}r Angewandte Mathematik,
Universit\"{a}t Bonn
(Germany)}
\email{albeverio@uni.bonn.de}
\address{
$^2$ \ SFB 611, \ Bonn, \ BiBoS, Bielefeld - Bonn\ (Germany)}
\address{
$^3$  CERFIM, Locarno and USI (Switzerland)}
\author{S.  N. Lakaev$^{4,6}$}
\address{
$^4$ Samarkand Division of Academy of sciences of Uzbekistan
(Uzbekistan)}
\email{lakaev@wiener.iam.uni-bonn.de}
\author{K. A. Makarov$^{5}$}
\address{
$^5$ Department of Mathematics, University of Missouri, Columbia,
MO, (USA)}
\email{makarov@math.missouri.edu}
\author{Z. I. Muminov$^6$}
\address{
$^6$ Samarkand State University, Samarkand (Uzbekistan)}
\email{zimuminov@mail.ru}



\begin{abstract}
For a wide class of  two-body energy operators  $h(k)$ on the
three-dimensional lattice $\bbZ^3$, $k$
 being the two-particle
 quasi-momentum, we prove that if the following two
assumptions (i) and (ii) are satisfied,
 then for all nontrivial values  $k$, $k\ne 0$,
 the discrete spectrum of  $h(k)$
 below its threshold  is non-empty. The assumptions are:
 (i) the two-particle
Hamiltonian $h(0)$ corresponding to the zero value of the
quasi-momentum has either an eigenvalue or a virtual level at the
bottom of its essential spectrum and (ii) the one-particle free
 Hamiltonians in the coordinate
representation generate  positivity preserving semi-groups.
\end{abstract}
 \maketitle

\section{Introduction}

The main goal of the present paper is to give a thorough
mathematical treatment of the spectral properties for the
two-particle lattice Hamiltonians with emphasis on {\it new
threshold phenomena}  that are not present in the continuous case
(see, e.g., \cite{AlbLakzM}, \cite{FIC}, \cite{Lak1}--\cite{Mat},
\cite{Mog} for relevant discussions and \cite{GS}, \cite{KM},
\cite{MS},
 \cite{Zh}
for the general
study of the low-lying excitation spectrum for  quantum
  systems on lattices).

 The kinematics of quantum quasi-particles
on lattices, even in the two-particle sector, is rather exotic.
For instance,
 due to the fact that   the discrete analogue of
the Laplacian or its generalizations (see \eqref{EE1} and \eqref{free}) are not
rotationally invariant,  the Hamiltonian of a system  does not separate into two parts, one
relating to the center-of-mass motion and the other one to the internal degrees
of freedom. In particular,  such a handy characteristics of inertia as
mass is not available. Moreover, such a natural local substituter
as the effective mass-tensor (of a ground state) depends on
the quasi-momentum of the system and, in addition, it  is only
semi-additive
 (with respect to the partial order on the set of positive
definite matrices). This is the so-called
{\it excess mass}
 phenomenon for lattice systems
 (see, e.g., \cite{Mat} and  \cite{Mog}):
 the effective mass of the bound state
of an $N$-particle system is greater than (but, in general,
 not equal to) the sum of the effective
masses of the constituent quasi-particles.

The two-particle  problem on lattices, in contrast to the continuous case where
the usual split-off of the center of mass can be performed,
can  be reduced to an effective   one-particle problem
 by using the Gelfand transform instead: the underlying
Hilbert space $\ell^2((\Z^3)^2)$
 is
decomposed as a direct von Neumann integral associated with the
representation of the discrete group $\Z^3$  by shift operators on the
lattice and then, the total two-body Hamiltonian appears to be
decomposable as well. In contrast to the continuous case, the
corresponding
  fiber Hamiltonians $h(k) $ associated with  the direct decomposition  depend
parametrically on the internal binding $k$, the quasi-momentum,
which  ranges over a cell of the dual lattice. As a consequence,
 due to the loss of the
spherical symmetry of the problem,  the spectra of the family
$h(k)$ turn out to be  rather sensitive to the variation of the quasi-momentum
$k$.

We recall that in the case of continuous Schr\"odinger operators
 one observes the emission of negative bound states from the
continuous spectrum at so-called critical potential strength (see,
e.g., \cite{AGH},  \cite{Lak2}, \cite{Rauch}, \cite{Yaf1}). This
phenomenon is closely related to the existence of
generalized eigenfunctions, which
are
 solutions of the Schr\"odinger equation
with zero energy decreasing at infinity, but are not square
integrable. These solutions are usually called 
 zero-energy resonance functions and, in
this case, the Hamiltonian is  called a critical one and the
Schr\"odinger operator is said to have a  zero-energy resonance (virtual level).
The appearance of negative bound states for critical
(non-negative) Schr\"odinger operators under infinitesimally small
negative perturbations is especially remarkable: it is the
presence of  zero-energy
resonances in at least two of the two-particle subsystems that
leads to the existence of
 infinitely many bound states for the corresponding three-body
system, the Efimov effect
  (see, e.g., \cite{AHW},  \cite{Lak1},\cite{OS},\cite{Sob},
 \cite{Tam1}, \cite{Tam2}  and \cite{Yaf2}).

It turns out that in the two-body  lattice case  there exists  an extra
mechanism for the bound state(s) to emerge  from the threshold of
the critical  Hamiltonians which has nothing to do with additional
(effectively negative) perturbations of the potential term. The
role of the latter is rather played by the adequate change of the
kinetic term which is due to the  nontrivial dependence of the
fiber Hamiltonians $h(k)$ on the quasi-momentum $k$ and
is related to the excess mass
 phenomenon for lattice systems mentioned above.

The main result of the paper, Theorem \ref {main}, is the
(variational) proof  of
 existence of the discrete spectrum  below
the bottom of the essential spectrum of the fiber
Hamiltonians $h(k)$  for
all non-zero values of the quasi-momentum $0\ne k \in
\T^3,$ provided that  the Hamiltonian
 $h(0)$  has either a virtual level or a
 threshold eigenvalue.

Apart from some technical smoothness assumptions upon  the {\it
dispersion relation of normal modes} $\varepsilon_\alpha(p), $
characterizing the free particles
$\alpha=1,2$,
  and as well as on smoothness assumptions (in the momentum representation)
  on the
two-particle interactions (Hypothesis  \ref{hypo}) the only
additional assumption   made (Hypothesis \ref{crat}) is that the one-particle free Hamiltonians
(in the coordinate representation)
 generate  {\it positivity preserving} semi-groups
$\exp(-t\hat h_\alpha^0),$ $ t> 0,\quad \alpha=1,2$. We remark
that this property is automatically fulfilled for the standard
Laplacian (discrete or continuous).

 The paper is
organized as follows.

In Section 2 we formulate the main hypotheses on the one-particle
lattice systems and prove  the basic inequality (see Lemma
\ref{neraven11} below) for the dispersion relations that are
conditionally negative definite. In Section 3 we introduce the
concept of a virtual level for the lattice one-particle
Hamiltonians and develop the necessary background for our further
considerations. In Section 4 we describe  the two-particle
Hamiltonians  in both the coordinate and the momentum
representation, introduce the two-particle quasi-momentum, and
decompose the energy operator into the von Neumann direct integral
of the  fiber Hamiltonians $h(k)$, thus providing the reduction to
the effective one-particle case.

In Section 5 we obtain efficient bounds on the location of the
discrete spectrum for the two-particle fiber Hamiltonian and  prove
the main result of this  paper,
Theorem \ref{main}, in the case where $h(0)$ has either a threshold
eigenvalue or virtual level at the bottom of its essential spectrum.

In Appendix \ref{Appendix I}, for readers convenience, we give a proof of
Proposition \ref{ker} which is a ``lattice'' analogue of a result due
to Yafaev \cite{Yaf3} in the continuous case.

In Appendix \ref{Appendix II} we construct an explicit example of a
one-particle discrete Schr\"odinger operator
 on the three-dimensional lattice $\Z^3$
that possesses both a virtual level  and threshold eigenvalue
 at the bottom of its essential spectrum (cf., e.g.,  \cite{AGH}, \cite{AlGHH}, and \cite{KSh} for related
discussions in the case of continuous Schr\"odinger operators).

\section{The one-particle Hamiltonian}
\subsection{Dispersion relations.}
 The free Hamiltonian
$\hat h^0$ of a
  quantum particle   on the  three-dimensional
 lattice $\bbZ^3$ is usually associated with the following self-adjoint
(bounded) multidimensional
  Toeplitz-type
 operator on the Hilbert space
$\ell^2(\bbZ^3)$ (see, e.g., \cite{Mat}):
\begin{equation}\label{EE1}
( \hat h^0\hat{\psi})(x)= \sum_{s\in {\Z}^3}\hat{\varepsilon}(s)
\hat{\psi}(x+s),\quad \hat{\psi} \in \ell^2(\Z^3).
\end{equation}
Here the series $\sum_{s\in \Z^3}\hat{\varepsilon} (s )$ is
assumed to be absolutely convergent, that is, 
\begin{equation*}
\{\hat{\varepsilon} (s )\}_{s\in \Z^3}\in \ell^1(\Z^3).
\end{equation*}
We also assume that  the ``self-adjointness'' property is
fulfilled 
 \begin{equation*}
 \hat{\varepsilon} (s )=\overline{\hat{\varepsilon} (-s )},\quad s\in \Z^3.
\end{equation*}

In the physical literature, the symbol of the Toeplitz operator
$\hat h^0$ given by the Fourier series 
\begin{equation*}
\varepsilon(p) =\sum_{s\in \Z^3} \hat
\varepsilon(s)e^{\mathrm{i}(p,s)},\quad p \in \T^3,
\end{equation*}
 being  a real valued-function on $\T^3$,   is called the {\it
dispersion relations of normal modes} associated with  the free
particle in question. The one-particle free Hamiltonian is
required to be of the form
$$
\hat h^0=\varepsilon(-\mathrm{i}\nabla),
$$
 where $\nabla$  is the generator of
the infinitesimal translations.

 Under the mild assumption that $$
 \hat v\in \ell^\infty(\Z^3),$$
where $\hat v=\{\hat v(s)\}_{s\in \Z^3}$ is a sequence of reals,
 the one-particle Hamiltonian $\hat h$,
  $$
 \hat h=\hat h^0+\hat v,
 $$
 describing the quantum particle moving
 in the potential field
 $\hat v$,
 is
 a bounded self-adjoint operator on the Hilbert space
$\ell^2(\Z^3)$.

 The one-particle Hamiltonian $h$  in the momentum
representation is introduced as
$$
h=\cF^{-1}\hat h\cF,
$$
where $\cF$ stands for the standard Fourier transform
    $\cF:L^2(\T^3) \longrightarrow   \ell^2(\Z^3)$,
 and
 $\T^3$ denotes  the three-dimensional torus, the
cube $(-\pi,\pi]^3$ with appropriately  identified sides.
Throughout the paper the torus   $\T^3$
 will always be considered as an
abelian group with respect to the addition and
multiplication by real numbers
 regarded as operations
on $\R^3$ modulo $(2\pi \Z)^3$.

 \subsection{Hamiltonians generating the  positivity preserving semi-groups}
The following  important  subclass of the  one-particle systems
  is of certain interest (see, e.g., \cite{CL}).
It is introduced by the additional requirement  that
 the dispersion relation $\varepsilon(p)$
is a real-valued
continuous
  conditionally negative definite function.
  Recall (see, e.g., \cite{RSIV}) that a complex-valued bounded
function $\varepsilon:\T^m\longrightarrow \R$ is called
conditionally negative definite if
$\varepsilon(p)=\overline{\varepsilon(-p)}$  and
\begin{equation}\label{nn}
  \sum_{i,j=1}^{n}\varepsilon(p_i-p_j)z_i\bar z_j\le 0
 \end{equation}
for all  $p_1, p_2,
.., p_n\in \T^m$ and all ${\bf z}=(z_1, z_2, ..., z_n)\in \C^n$
satisfying $\sum_{i=1}^nz_i=0$.

  It is known that in this case
  the dispersion relation $\varepsilon(p)$
   admits the (L\'evy-Khinchin) representation
  (see, e.g., \cite{BCR})
  $$
  \varepsilon(p)=\varepsilon(0)+\sum_{s\in \Z^3\setminus\{0\}}(e^{\mathrm{i}(p,s)}-1)\hat
  \varepsilon (s),\quad p\in \T^3,
  $$
  which is equivalent to the requirement that  the Fourier
coefficients $\hat \varepsilon(s)$ with $s\ne 0$   are
non-positive, that is, 
\begin{equation*}
 \hat \varepsilon(s)\le 0, \quad s\ne 0,
\end{equation*}
and the series $\sum_{s\in \Z^3\setminus\{0\}}\hat
  \varepsilon (s)$ converges absolutely.
In turn, this is also equivalent
  to  that the
lattice Hamiltonian $\hat h=\hat h^0+v$ generates
 the positivity preserving
 semi-group $ e^{-t\hat h} $, $t>0,$  on $\ell^2(\Z^3)$
(see, e.g., \cite{RSIV} Ch. XIII). Following \cite{CL} we call the
free Hamiltonians $\hat
h^0=\varepsilon(- \mathrm{i}{\bf \nabla})$ generating the
positivity preserving semi-groups
 {\it  the generalized Laplacians}.

The following example shows that the standard discrete Laplacian
is a generalized Laplacian in the sense mentioned above.

\begin{example}\label{ex}
   For   the one-particle   free Hamiltonian
\begin{equation*}
(\hat h^0\hat{\psi})(x)= (-\Delta\hat{\psi})(x)= \sum_{\mid s\mid
=1} [\hat{\psi}(x)- \hat{\psi}(x+s)],\quad x\in \Z^3, \quad
\hat{\psi}\in\ell^2({\Z}^3),
\end{equation*}
the (Fourier) coefficients $\hat \varepsilon(s)$, $s\in \Z^3$,
 from \eqref{EE1}
 are necessarily of the form
\begin{equation*}
\hat{\varepsilon} (s)=
 \begin{cases}
6, & s=0\\
-1, & |s|=1\\
0, & \text{otherwise}.
 \end{cases}
 \end{equation*}
Hence, the corresponding dispersion relation
\begin{equation}\label{disdis}
\varepsilon(p)= 2\sum_{i=1}^{3}(1-\cos p_i),\quad p=(p_1, p_2,
p_3)\in \T^3,
 \end{equation}
is a conditionally negative definite function.
\end{example}

 We  need a simple inequality which will play a crucial role in the proof
 of the main results of the paper, Theorems  \ref{main0} and  \ref{main}.

\begin{lemma}\label{neraven11}
Assume that  the dispersion relation $\varepsilon(p) $  is a real-valued continuous
conditionally negative definite function on $\T^3$. Assume, in
addition, that $\varepsilon(0)$ is the unique minimum of the
function
$\varepsilon(p)$. Then for all $q\in \T^3\setminus\{0\}$ the
inequality
\begin{equation}\label{in}
\varepsilon (p)+\varepsilon (q)
>\frac{\varepsilon (p+q)+\varepsilon (p-q)}{2} +\varepsilon(0)
, \quad \text{a.e.} \quad p\in {\bbT}^3,
\end{equation}
holds.
\end{lemma}

\begin{proof}
For $n=4$, $m=3$,
\begin{align*}
p_1&=\frac{p+q}{2},
\quad
p_2=\frac{q-p}{2},
\\ \no
p_3&=-p_1,
 \quad
\quad p_4=-p_2,
 \end{align*}
and
 $
{\bf z}=( 1, -1, 1, -1) $ applying  inequality \eqref{nn}
($\varepsilon(p)$ is an even function) proves the non-strict
version of inequality \eqref{in}. To show that inequality
\eqref{in} is strict we proceed as follows.

Fix a $q\in \T^3$, $q\ne 0$. Then there exists an $s_0\in
\Z^3\setminus \{0\}$ such that $\hat{\varepsilon}(s_0)<0$ and
$\cos (q,s_0)\ne 1$ (otherwise $\varepsilon(q)=\sum_{s\in\Z^3}
\hat \varepsilon (s)=\varepsilon (0) $ which contradicts the
hypothesis that $\varepsilon(0)$ is the unique minimum of the
function $\varepsilon(\cdot)$ on $\T^3$). Since the cosine
function is conditionally positive definite, using the non-strict
version of inequality \eqref{in} yields
\begin{align}
F(p,q)&\equiv\varepsilon (p)+\varepsilon (q)
-\frac{\varepsilon (p+q)+\varepsilon (p-q)}{2} -\varepsilon(0)
\no \\
& =\sum_{s\in\Z^3} \hat \varepsilon(s) \bigg [\cos
(p,s)+\cos (q,s)-\frac{\cos (p+q,s)+\cos (p-q,s)}{2}-1\bigg
]
\no \\
&\ge 2 \hat
\varepsilon(s_0) \bigg [\cos (p,s_0)+\cos (q,s_0)-\frac{\cos (p+q,s_0)+\cos
(p-q,s_0)}{2}-1\bigg ]
\no \\
&= 2 \hat \varepsilon(s_0) \bigg [\cos (p,s_0)+\cos
(q,s_0)-\cos (p  ,s_0)\cos (q,s_0)-1\bigg ]\no.
\end{align}
 Hence
$$F(p,q)\ge 2 \hat
\varepsilon(s_0) \bigg [(\cos (p,s_0)-1)(1-\cos (q,s_0))\bigg ]>0,
\quad (p,s_0)\ne 2n\pi, \quad n\in \Z,
$$
 since $\hat\varepsilon
(s_0)<0$ and $\cos(q, s_0)\ne 1$, completing the proof.
\end{proof}

 \section{A virtual level  and  threshold eigenvalues}
In order to  introduce the concept of a {\it virtual level} ({\it
threshold resonance})
 for the (lattice) energy
operator $ h$ we assume the following technical hypotheses that
guarantee some smoothness of the dispersion relation
$\varepsilon(p)$ and the continuity of the Fourier transform
$$
v(p)=(2\pi)^{-\frac{3}{2}}\sum_{s\in \Z^3}\hat
v(s)e^{\mathrm{i}(p,s)}
$$ of the interaction $\hat v$.

 \begin{hypothesis}\label{hypo}  Assume that the dispersion relation
$\varepsilon(p)$   is a twice differentiable (periodic)
real-valued functions on $\T^3$ with a unique non-degenerate
minimum at the origin. Assume, in addition, that $v(p)$  is a
continuous function on   $\T^3$ such that
 $$
v(p)=\overline{v(-p)}, \quad p\in \T^3. $$
 \end{hypothesis}

For $\lambda\le
\varepsilon(0)$ on the Banach space $C(\T^3)$ of continuous
(periodic) functions on $\T^3$ we shall consider
 the integral operator $G(\lambda)$  with the (Birman-Schwinger)
kernel function
\begin{equation}\label{GG}
G(p,q;\lambda)=(2\pi)^{-\frac{3}{2}} v(p-q)
(\varepsilon(q)-\lambda)^{-1}, \quad
p,q\in \T^3.
\end{equation}

\begin{lemma}\label{comp}
Assume Hypothesis  \ref{hypo}. Then
for $\lambda\le \varepsilon(0)$ the operator $G(\lambda)$ on $C(\T^3)$
given by \eqref{GG}
is compact.

\end{lemma}
\begin{proof}
Given  $f\in L^1(\T^3)$, for  the function $g$  introduced by
$$
g(p)=(2\pi)^{-\frac{3}{2}} \int\limits_{{\T}^3} v
(p-q)f(q)dq
$$
 one has the estimates
\begin{equation}\label{ac1}
|g(p)|
\le(2\pi)^{-\frac{3}{2}}\sup_{p, q\in \T^3}|v
(p-q)| \|f\|_{L^1(\T^3)}
\end{equation}
and
\begin{equation}\label{ac2}
|g(p+\ell)-g(p)|=\bigg |(2\pi)^{-\frac{3}{2}} \int\limits_{{\T}^3} (v
(p+\ell-q)-v(p-q))f(q)dq\bigg |
\end{equation}
\begin{equation*}
\le(2\pi)^{-\frac{3}{2}}\sup_{t\in \T^3}|v (t+\ell)-v(t)|
\|f\|_{L^1(\T^3)}.
\end{equation*}
 Since for  $\lambda\le
\varepsilon(0)$, the function $(\varepsilon(\cdot)-\lambda)^{-1}$,
is summable, the multiplication
operator by the function $(\varepsilon(\cdot)-\lambda)^{-1}$  from
$C(\T^3)$ into $L^1(\T^3)$ is continuous. Therefore, from
\eqref{ac1} and \eqref{ac2} it follows that the image of the unit
ball in $C(\T^3)$ consists of functions that are totally  bounded
and equicontinuous: $v$ is continuous and,
 therefore,
$$\lim_{|\ell|\to 0}\sup_{t\in \T^3}|v (t+\ell)-v(t)|
\|f\|_{L^1(\T^3)}=0.$$

An application of
 the Arzela-Ascoli Theorem  then completes the proof.
\end{proof}
\begin{remark}\label{virt}
Clearly (cf. \cite{Yaf3}), the operator $h$ has an
eigenvalue $\lambda\le \varepsilon(0)$, that is,
$\Ker (h-\lambda I)\ne 0$,
if and
only if  the compact
operator $G(\lambda)$ on $C(\T^3)$ has  an eigenvalue $-1$ and there exists a function
$\psi\in \Ker(G+I)$ such that
the function $f$ given by
$$
f(p)=\frac{\psi(p)}{\varepsilon(p)-\lambda}
 \quad \text{a.e.}\quad p\in \T^3,
$$
belongs to $L^2(\T^3)$. In this case $f\in \Ker(h-\lambda I)$.

Moreover, if $\lambda<\varepsilon(0)$, then
\begin{equation}\label{iadro}
\dim\Ker (h-\lambda I)=\dim \Ker (G(\lambda)+I)
\end{equation}
and
$$
\Ker (h-\lambda I)=\{f\,\vert\,
f(\cdot)=\frac{\psi(\cdot)}{\varepsilon(\cdot)-\lambda}\, , \, \psi \in
\Ker (G(\lambda)+I)\}.
$$
In the case of a threshold eigenvalue
$\lambda=\varepsilon(0)$ equality \eqref{iadro}
may fail to hold. It should be replaced by
the  inequality
\begin{equation*}
\dim\Ker (h-\varepsilon(0) I)\le\dim \Ker
(G(\varepsilon(0))+I).
\end{equation*}
\end{remark}

In order to discuss the threshold phenomena,
 that is,  the case $\lambda=\varepsilon(0)$,
following \cite{AlGHH} and \cite{JK}  (see also \cite{KSh} for a
related discussion), under Hypothesis \ref{hypo}
we distinguish five mutually disjoint cases:

{\it Case }I:  $-1$ is not an eigenvalue of
$G(\varepsilon(0))$.

{\it Case }II: $-1$ is  a simple  eigenvalue of $G(\varepsilon(0))$ and
 the
associated eigenfunction $\psi$ satisfies the condition
\begin{equation*}
\frac{\psi(\cdot)}{\varepsilon(\cdot)-\varepsilon(0)}\notin
L^2(\T^3).
\end{equation*}

{\it Case } III: $-1$ is  an eigenvalue of $G(\varepsilon(0))$ and any  of  the
associated eigenfunctions $\psi$ satisfies the condition
\begin{equation*}
\frac{\psi(\cdot)}{\varepsilon(\cdot)-\varepsilon(0)}\in
L^2(\T^3).
\end{equation*}

{\it Case }IV: $-1$ is  a multiple eigenvalue of $G(\varepsilon(0))$ and exactly
one (up to a normalization) of the associated eigenfunctions
$\psi$ satisfies the condition
\begin{equation*}
\frac{\psi(\cdot)}{\varepsilon(\cdot)-\varepsilon(0)}\notin
L^2(\T^3).
\end{equation*}

{\it Case }V: $-1$ is  a multiple eigenvalue of $G(\varepsilon(0))$ and at least
two   of the associated eigenfunctions $\psi$  and $\varphi$ that
are linearly independent satisfy the condition
\begin{equation*}
\frac{\psi(\cdot)}{\varepsilon(\cdot)-\varepsilon(0)}\notin
L^2(\T^3) \quad \text{and } \frac{\varphi
(\cdot)}{\varepsilon(\cdot)-\varepsilon(0)}\notin
L^2(\T^3).
\end{equation*}

Given the classification above, we arrive at the following definition of a
virtual level.

\begin{definition}\label{def} In Cases II, IV and V
 the operator $h$
is said to have a virtual level (at the threshold).
\end{definition}
\begin{remark} Note that in Cases III and IV the
operator $h$ has a threshold
eigenvalue $\lambda=\varepsilon(0)$ of
 multiplicity $\dim \Ker(G(\varepsilon(0))+I)$ and
 $\dim \Ker(G(\varepsilon(0))+I)-1$ respectively.
\end{remark}

\begin{remark}
Our definition  of a virtual level is
equivalent to the direct analogue of that
 in the continuous case
 (see, e.g.,  \cite{AGH}, \cite{Sob}, \cite{Tam2}, \cite{Yaf2},
 \cite{Yaf3} and references therein).
\end{remark}
\begin{remark}\label{l1l1}
If the Hamiltonian $h$  has a virtual level and
 the corresponding function $\psi$ , $\psi\in \Ker(G(\varepsilon(0))+I),$ is such that $
\frac{\psi(\cdot)}{\varepsilon(\cdot)-\varepsilon(0)}\notin
L^2(\T^3) $, then the
  function
\begin{equation}\label{fun}
f(p)=\frac{\psi(p)}{\varepsilon(p)-\varepsilon(0)}, \quad p\in \bbT^3,
\end{equation}
belongs to $ L^{r}(\bbT^3)$, $1\le r<3/2$.
 
In particular, the function $f$ given by \eqref{fun}  is the
eigenfunction of the operator $h$ associated with the eigenvalue
$\varepsilon(0)$ in the Banach space $L^1(\T^3)$, that is,
$$hf=\varepsilon(0)f$$
and hence the following equation 
\begin{equation*}
 \varepsilon (p)f(p)+(2\pi)^{-\frac{3}{2}}\int_{\T^3}
 v(p-q)f(q)dq=\varepsilon (0)f(p),
 \quad \text{ a.e. }  p\in \T^3,
 \end{equation*}
 holds.

A simple computation shows that the Fourier coefficients $\hat
f(s)$, $s\in \Z^3,$ of the (summable) function  $f$  solve the
infinite system of homogeneous equations 
\begin{equation*}
 \sum_{s\in
{\Z}^3}\hat{\varepsilon}(s) \hat{f}(x+s)+(\hat v(x)-\varepsilon(0))\hat{f}(x)=0,\quad x\in \Z^3,
\end{equation*}
and hence the equation (in the coordinate representation)
$$
\hat h\hat f=\varepsilon(0)\hat f
$$
has a  solution $\hat f$, a threshold resonant state, that does not belong to $\ell^2(\Z^3)$
but vanishes at infinity,
$$\lim_{|s|\to \infty}\hat f (s)=0
$$
(by the Riemann-Lebesgue Theorem).
\end{remark}
\begin{remark}\label{evemodd}
If  the dispersion relation $\varepsilon(p)$ is known to be  an even function,
$\varepsilon (p)=\varepsilon(-p),$ or, which is the same, the
Fourier coefficients satisfy the condition
$$
\hat \varepsilon(s)=\hat \varepsilon(-s)\in \R, \quad s\in \Z^3,
$$
the Birman-Schwinger kernel $G(p,q;\lambda)$ has the additional  property that
$$
G(p,q;\lambda)=\overline{G(-p, -q;\lambda)}.
$$
Hence, if $\psi\in \Ker(G(\lambda)+I)$, $\lambda\le
\varepsilon(0)$,
 so does the function
$\varphi(p)=\overline{\psi(-p)}$. Therefore, exactly one of the
functions $\psi\pm\varphi$ is also an eigenfunction of
$G(\lambda)$ associated with the eigenvalue $-1$, and hence,
without loss of generality one may assume that the operator
$G(\lambda)$ has an eigenfunction $\tilde \psi$ such that $|\tilde
\psi(\cdot)|$ is an even function.

\end{remark}

To get  finer results (cf. \cite{Yaf3}) we  need an auxiliary
scale of the Banach spaces $\cB (\mu)$, $0<\mu\le 1$, of H\"older
continuous functions on $\T^3$  obtained by the closure of the
space of smooth (periodic) functions $f$ on $\T^3$ with respect to
the  norm 
\begin{equation*}
\|f\|_\mu=\sup_{t, \ell\in \T^3}\bigg
[|f(t)|+|\ell|^{-\mu}|f(t+\ell)-f(t)|\bigg ].
\end{equation*}
Note that the spaces $\cB(\mu)$ are naturally embedded one into the other
$$
\cB(\nu)\subset \cB(\mu)\subset C(\T^3), \quad 0<\mu\le \nu\le 1.
$$

 If
 $v\in \cB(\kappa) $ with $\kappa >\frac{1}{2}$, the
  following proposition, a variant of the Birman-Schwinger
principle, is a convenient tool to decide whether the threshold
 $\varepsilon(0)$   of the essential spectrum of $h$
 is an eigenvalue   (resp. a virtual level ) for the operator $h$.

\begin{proposition}\label{ker}(cf. \cite{Yaf3}) Assume Hypotheses \ref{hypo}.
Assume, in addition, that $v\in \cB(\kappa) $
  with $\kappa>\frac{1}{2}$.
Then the operator $h(0)-\varepsilon(0)I$ has a non-trivial kernel  if and only if
$-1$ is  an eigenvalue of $G(\varepsilon(0))$ and
one of  the
associated eigenfunctions $\psi$ satisfies the condition
\begin{equation*}
\psi(0)=0.
\end{equation*}
In particular, the operator $h$ has a virtual level if and only if
$-1$ is  an eigenvalue of $G(\varepsilon(0))$ and
one of  the
associated eigenfunctions $\psi$ satisfies the condition
\begin{equation*}
\psi(0)\ne0.
\end{equation*}
\end{proposition}
\begin{proof} See Appendix \ref{Appendix I}.\end{proof}
\begin{remark}  A more thorough examination of the  proof  shows that, in fact,
$$
\dim \Ker (h(0)-\varepsilon(0)I)=\begin{cases}
\dim \Ker (G(\varepsilon(0))+I),& \text{in Case III}\\
\dim \Ker (G(\varepsilon(0))+I)-1,& \text{in Case IV}.
\end{cases}
$$

\end{remark}

\begin{remark}\label{rrr}

If $\kappa>\frac{1}{2}$, as it follows from Proposition \ref{ker},
the eigensubspace of functions
  $\psi$  associated with
the eigenvalue $\lambda=-1$  of $G(\varepsilon(0))$ with
  the additional constraint $\psi(0)\ne0$  is one-dimensional.
  This
   proves that {\it Case} V  does not occur if $\kappa >\frac{1}{2}$ .
In Case IV  we  have the coexistence of a (simple) virtual
level
  and
a (possibly multiple) threshold eigenvalue (see Appendix \ref{Appendix II} for a
concrete example of such a coexistence in {\it Case} IV).
\end{remark}
\begin{remark}\label{empty} It is known that
for the continuous   Schr\"odinger operators $h=-\Delta+V(x)$ with
  $V\in L^1(\R^3)\cap R$, $R$ the Rollnik class,  {\it Case } V does not occur
(see, \cite{AGH} Lemma 1.2.3). It is also worth mentioning that
if, in addition, the Schr\"odinger operator $h$ is non-negative, then
under the $L^{\frac{3}{2}}$-weak assumption on the potential $V$
  {\it Cases}  III,  and  IV
   do not occur (there is no zero-energy
  eigenstate)
  (see, e.g., \cite{KS}, \cite{S}, and \cite{Tam1}).
\end{remark}

\section{The  two-particle Hamiltonian. Reduction to the
one-particle case}
\subsection{ The coordinate representation}

  The free Hamiltonian $\widehat H^{0}$ of the system of
 two quantum particles $\alpha=1,2$,  with the dispersion relations
 $\varepsilon_\alpha(p), $ $ \alpha=1,2,$ respectively, is introduced
(as a bounded self-adjoint operator on the Hilbert space
$\ell^2((\bbZ^3)^2)\simeq \ell^2(\bbZ^3)\otimes\ell^2(\bbZ^3)$) by
\begin{equation}\label{free}
\widehat{H}^{0}=\hat h_1^{0}\otimes I+I \otimes \hat h_2^{0},
\end{equation}
with
$$
\hat h^{0}_\alpha=\varepsilon_\alpha(-\mathrm{i}\nabla), \quad
\alpha=1,2,
$$
and  $I$ the  identity operator on
$\ell^2(\Z^3)$.

The total  Hamiltonian  $\widehat H$ (in the coordinate
representation) of the two-particle system  with the real-valued
pair interaction $\widehat V$ is a self-adjoint bounded operator on
the Hilbert space $\ell ^2(({\Z}^3)^2)$ of the form
\begin{equation*}
\widehat H =\widehat H^{ 0}+ \widehat{V},
\end{equation*}
where
\begin{equation*}
(\widehat{V}\hat{\psi})(x_1,x_2) ={\hat{v}(x_1 -x_2
)\hat{\psi}(x_1,x_2)},\quad \hat \psi \in \ell ^2(({\Z}^3)^2),
\end{equation*}
with $\{\hat v(s)\}_{s\in \Z^3}$ the Fourier coefficients of a
continuous function $v(p)$  satisfying  Hypothesis \ref{hypo}.

 \subsection{ The momentum  representation}

The transition to the momentum representation is performed by the
standard Fourier transform
  $${ \cF}_2:L^2(({\T}^3)^2) \longrightarrow \ell^2((%
{\Z}^3)^2),$$ 
 where ${({\T}^3)^m}$ denotes the Cartesian
$m$-th power of the three-dimensional cube  ${\T}^3=(-\pi,\pi]^3:$
\begin{equation}\label{ft}
({\T}^3)^m=\underbrace{{\T}^3\times {\T}^3 \times \cdots \times
{\T}^3 }_ {m \quad \mbox{times}} ,\quad m\in {\N}.
\end{equation}

 The two-particle
Hamiltonian $H$ in the momentum representation is then given  by
\begin{equation*}
H =H^0+V,
 \end{equation*}
 where
\begin{equation*}
(H^0f)(k_1 ,k_2 )=(\varepsilon_1(k_1)+ \varepsilon _2 (k_2 ))f(k_1
,k_2 ) ,\quad f \in L^2(({\T}^3)^2),
\end{equation*}
 and  $V$ is the operator of
partial integration given by
\begin{equation*}
(V f)(k_1 ,k_2 )=(2\pi)^{-\frac{3}{2}}{\int\limits_{({\T}^3)^2}} v
(k_1-k_1' ) \delta (k_1 +k_2 -k_1 '-k_2 ')f(k_1 ',k_2')dk_1
'dk_2',
 \end{equation*}
\begin{equation*}
f\in L^2(({\T}^3)^2).
\end{equation*}
Here the kernel function is given by the Fourier series
\begin{equation*}
v (p)=(2\pi )^{-3/2}\sum_{s\in {{{\Z^3}}}}  \hat v
(s)\,e^{\mathrm{i}(p,s)},\quad p \in \T^3,
\end{equation*}
 and $\delta (p)$
 denotes the  Dirac
delta-function.

\subsection{Direct integral decompositions. The quasi-momentum.}
Denote by $\hat U^2_s$, $s\in \Z^3$, the unitary  representation of
the
 abelian group  $\Z^3$ by the shift operators on the Hilbert
 space $\ell^2(({\Z}^3)^2)$:
\begin{equation*}
(\hat U^2_s \hat \psi )(n_1,n_2)=\hat \psi (n_1+s,n_2+s),\quad
\hat \psi \in \ell^2((\Z^3)^2),\quad n_1, n_2, s\in \Z^3.
\end{equation*}

Via the Fourier transform $ \cF_2$  the unitary representation
\begin{equation*}
  \hat U^2_{s+t}=\hat U^2_s\hat U^2_t,\quad s,t \in \Z^3,
  \end{equation*}
 induces the representation of the
group $\Z^3$ in the Hilbert space $L^2(({\T}^3)^2)$  by unitary
(multiplication) operators $U^2_s= \cF_2^{-1}\hat U^2_s\cF_2$,
$s\in \Z^3$,
\begin{equation}\label{grup}
(U_s^2 f)(k_1,k_2)= \exp \big (
-\mathrm{i}(s,k_1+k_2)\big)f(k_1,k_2),\quad k_1, k_2\in
\T^3, \quad f\in L^2((\T^3)^2).
\end{equation}

Given   $k\in {\T}^3$, we define $\F_k$ as follows
\begin{equation*}
\F_k=\{(k_1,k-k_1){\ \in }({\T}^3)^2: k_1\in \T^3,\,\,k-k_1 \in
\T^3\}.
\end{equation*}
Introducing the mapping
\begin{equation*}
\pi:(\T^3)^2\to \T^3,\quad
\pi ((k_1, k_2))=k_1,
\end{equation*}
we denote by  $\pi_k$, $k\in \T^3$, the restriction of $\pi $ to
$\F_k\subset (\T^3)^2$, that is,
\begin{equation}\label{project}
\pi_{k}= \pi \vert_{\F_k}.
\end{equation}
We remark  that $ \F_k $, $k\in {\T}^3$, is a three-dimensional
manifold homeomorphic to
 ${\bbT}^3$.

The following lemma is evident.
\begin{lemma}
 The mapping  $\pi_{k}$, $k\in \T^3$,
from $\F_k\subset (\T^3)^2$ onto $\T^3$ is bijective, with the inverse mapping
given by
\begin{equation*}
(\pi_{k})^{-1}(q )=(q ,k-q ).
\end{equation*}
\end{lemma}

Decomposing the Hilbert space $ L^2((\T^3)^2)$  into the direct
integral
\begin{equation*}
L^2((\T^3)^2)= \int_{k\in {\T}^3} \oplus L^2(\F_k)dk
\end{equation*}
yields the corresponding decomposition of the unitary
representation  $U_s^2$,
 $s \in \Z^3$, into the  direct integral
\begin{equation*}
U_s^2= \int_{k\in {\T}^3} \oplus U_s(k)dk,
\end{equation*}
with
\begin{equation*}
U_s(k)=e^{-\mathrm{i}(s,k)}I_{L^2(\F_k)}
\end{equation*}
and $I_{L^2(\F_k)}$  the identity operator on the Hilbert
space $ L^2(\F_k)$.

  The
Hamiltonian  $\widehat H$ (in
the coordinate representation) obviously commutes with the  group
of translations, $\hat U^2_s$, $s\in \Z^3$,
 that is,
\begin{equation*}
\hat U^2_s\widehat H=\widehat H\hat  U^2_s, \quad s\in \Z^3.
\end{equation*}
So does the  Hamiltonian  $ H$ (in the momentum representation)
with respect to the  group  $ U^2_s$, $s\in \Z^3$, given by
\eqref{grup}.
  Hence, the operator
$H$ can be decomposed into the direct integral
 \begin{equation}\label{dirint}
H= \int_{k \in {\T}^3}\oplus\tilde h(k)dk
 \end{equation}
 associated with the decomposition
\begin{equation*}
L^2 (( {\T}^3)^2) =\int_{k \in {\T}^3} {\ \oplus } L^2 ( \F_k )
dk.
\end{equation*}
In the physical literature the parameter $k$, $k\in \T^3$, is
called the {\it two-particle  quasi-momen\-tum} and the corresponding operators
 $\tilde h(k),$ \,$k \in {\T}^3,$ are called the
{\it fiber operators}.

\subsection{The two-particle dispersion relations.}

 The fiber operators $\tilde h(k),$ \,$k \in {\T}^3,$
from the decomposition \eqref{dirint} are unitarily equivalent to
the operators
 $h (k)$, $ k \in {\T}^3$,
of the form 
\begin{equation*}
 h (k) =h^{0} (k)+v,
\end{equation*}
 where
\begin{equation*}
(h^{0}(k)f)(p)=\cE_k(p)f(p),
\end{equation*}
 \begin{equation*}
 (v
f)(p)=(2\pi)^{-\frac{3}{2}} \int\limits_{{\T}^3} v (p-q)f(q)d
q, \quad f \in L^2({\T}^3)
\end{equation*}
and the {\it two-particle dispersion relations}
 \begin{equation*}
 \cE_{k}(p)=
\varepsilon_{1} (p) +\varepsilon_{2} (k-p), \qquad p\in \T^3,
\end{equation*}
parametrically depend on the quasi-momentum $k $, $k\in \T^3$.

The   equivalence is given by the unitary operator
$
u_{k}:L^2(\F_{k}) \rightarrow L^2({\bbT}^3),\,\, k\in {\bbT}^3,
$
$$\,\, u_{k} g=g \circ (\pi_{k})^{-1},
$$
with $\pi_{k}$  defined by \eqref{project}.

\section{Spectral properties of the fiber operators $h(k)$ }
As we have learned from   the previous section, the two-particle Hamiltonian $H$ (up to unitary equivalence)
can be decomposed into the direct integral
$$
H\simeq  \int\limits_{k \in {\T}^3}\oplus h(k)dk,
$$
where the fiber operators $$h(k)=h^0(k)+v$$ can be considered as
 the one-particle Hamiltonians with the {\it two-particle dispersion relations}
\begin{equation}\label{kinetik}
 \cE_{k}(p)=
\varepsilon_{1} (p) +\varepsilon_{2} (k-p), \qquad p\in \T^3,
\end{equation}
 with $\varepsilon_\alpha(p)$ the dispersion relations for the
 particles $\alpha=1,2.$

Under Hypothesis \ref{hypo} the perturbation $v$ of the operator
$h^0(k)$, $k\in\T^3$,  is a Hilbert-Schmidt operator
and, therefore, in accordance with the Weyl Theorem
 the essential
  spectrum of the operator $h(k)$ fills in the
following interval on the real axis:
$$
\sigma_{\text{ess}}(h(k))= [{ \cE}_{\min}(k) ,{\cE }_{\max}(k)],
$$
where
 $$
{\cE }_{\min}(k)=\min_{q{\in }{\bbT}^3}\cE_{k} (q),\quad {\cE }_{\max}
(k)=\max_{q{\in }{\bbT}^3} \cE_{k} (q).
 $$

 If the dispersion relations in the
 one-particle sector are conditionally negative definite, then so is the
 two-particle
 dispersion relation
 $\cE_0(p)$ corresponding to the zero-value of the quasi-momentum $k$.
 Hence, under these assumptions,  the Hamiltonian $h(0)$
 in the coordinate
 representation generates the positivity preserving semi-group
 $e^{-t h(0)}$, $t>0$
  (which is not necessarily true for the fiber  Hamiltonians
 $h(k)$
  with $k\ne 0$: the function
 $\cE_{k}(p)$ may not be even, and hence, not conditionally negative definite).
Although the two-particle dispersion relations are not necessarily
conditionally negative definite for nontrivial values of the quasi-momentum,
 they still satisfy some useful  inequality, Lemma
 \ref{twoner} below,  analogous to that
 in Lemma \ref{neraven11} for the one-particle dispersion relations.
\begin{hypothesis}\label{crat}
Assume Hypothesis \ref{hypo}.
 Suppose
that the dispersion relations $\varepsilon_\alpha (p)$, $\alpha=1,2$, in the
 one-particle sectors are conditionally negative definite.
\end{hypothesis}

\begin{lemma}\label{twoner}
Assume Hypothesis \ref{crat}.
 Then
for any (fixed) $k, q\in \T^3$ such that either $k\ne q$ or $q\ne 0$
 \begin{equation*}
 \cE_0(p)-\cE_{0}(0)+\cE_{k}(q)-
  \frac{ \cE_k(p+q)+
\cE_k(q-p)}{2}>0, \quad \text{a.e.}\quad p\in \T^3.
\end{equation*}
  In particular, if $k\ne 0$, and
   $p(k)$ is  a (any)  point where the function
   $\cE_k(\cdot)$ attains its minimal value, that is,
  $$
  \cE_{\min}(k)=\cE_k(p(k)),
  $$ the following inequality
  \begin{equation*}
  \cE_0(p)-\cE_{\min}(0) +\cE_{\min}(k)-
  \frac{ \cE_k(p+p(k))+
\cE_k(p(k)-p)}{2}>0,  \quad \text{a.e.} \quad p\in \T^3,
\end{equation*}
   holds.
\end{lemma}

\begin{proof}
Since $|q|^2+|k-q|^2\ne 0$ the claim is an immediate consequence of
  Lemma  \ref{neraven11} and definition \eqref{kinetik} of the
  two-particle dispersion relations:

\begin{align*}
& \cE_0(p)+\cE_k(q)-
  \frac{ \cE_k(p+q)+
\cE_k(q-p)}{2}-\cE_{0}(0)
\no \\
&= \varepsilon_1 (p)+\varepsilon_1 (q)
-\frac{\varepsilon_1 (p+q) +\varepsilon_1 (q
-p)}{2}-\varepsilon_1(0)
\\ \no
&+  \varepsilon_2 (p)+\varepsilon_2 (k -q) -\frac{\varepsilon_2
(k-q-p) +\varepsilon_2 (k-q+p)}{2}
-\varepsilon_2(0)>0,\quad \text{a.e.}\quad p\in \T^3.
\end{align*}
\end{proof}

Our first non-perturbative
result shows that under Hypothesis \ref{crat} the discrete
spectrum of the fiber operators $h(k)$
under the variation of the quasi-momentum cannot be absorbed by the
threshold, provided
that $h(0)$ has eigenvalues below the bottom of its essential spectrum.

\begin{theorem}\label{main0} Assume Hypothesis \ref{crat}.
 Denote by $m(k)$, $k\in \T^3$, the  lower bound of the
 operator $h(k)$,
 $$
 m(k)=\inf \spec ( h(k)), \quad k\in \T^3.
 $$
 If 
 the Hamiltonian $h(0)$
 does not  have
the discrete spectrum below the bottom of its essential spectrum, that is,
 $m(0)=\cE_{\min}(0)$, assume, in addition, that  the lower edge $m(0)$
  of the spectrum of the
  operator $h(0)$ is an eigenvalue.
  Then
  \begin{equation}\label{disdisdis}
  \cE_{\min}(0) - m(0)<\cE_{\min}(k) -m(k)
 , \quad k\in \T^3,
  \,\,k\ne 0.
  \end{equation}
 \end{theorem}

\begin{proof}  Let $0\ne f\in \Ker (h(0)-m(0)I)$ and hence
$$
 {\cE}_0(p)f(p)+(2\pi)^{-\frac{3}{2}}\int_{\T^3}
 v(p-q)f(q)dq=m(0)f(p),
 \quad \text{ a.e. }  p\in \T^3.
 $$

  By hypothesis the one-particle
 dispersion relations are conditionally negative definite functions.
Then, as it can easily be seen from the definition of the two-body
 dispersion relation, the function
 $\cE_0(p)$ corresponding to the zero-value of the quasi-momentum $k$ is
 also conditionally negative definite. In particular,
 $\cE_0(p)$ is an even functions
 and, hence, by Remark
 \ref{evemodd},
 without loss of
 generality one may assume that the function $|f(\cdot)|$  is even.

 For  $k\in \bbT^3$ we introduce  the trial $L^2(\T^3)$-function
 \begin{equation*}
f_{k}(p)= f(p-p(k)),
\end{equation*}
 where $p (k)$ denotes  the minimum point of the function
$\cE_{{k}}(p)$, that is,
$
\cE_{{k}} (p (k))=\cE_{\text{min}}(k)
$
(if the minimum value of $\cE_{k}(p)$ is attained in several points
 choose $p(k)$ as any  one of them arbitrarily).

To prove \eqref{disdisdis}
 it is sufficient to
establish the inequality
\begin{equation}\label{dno}
\Gamma(k)=([h(k)- (\cE_{\text{min}}(k)-\cE_{\text{min}}(0)+m(0))]f_{k},f_{k})< 0, \quad k\ne
0.
\end{equation}
One gets
\begin{align}
&([h(k)-(\cE_{\text{min}}(k)-\cE_{\text{min}}(0)+m(0))]f_k)(p)
\label{nol}\\
&=[\cE_k(p)-(\cE_{\text{min}}(k)-\cE_{\text{min}}(0)+m(0))]f(p-p(k))
\no \\
&+
(2\pi)^{-\frac{3}{2}} \int_{\T^3} v(p-q)f(q-p(k))dq
\no\\
&=
[\cE_k(p)-(\cE_{\text{min}}(k)-\cE_{\text{min}}(0)+m(0))]f(p-p(k))
\no \\
&+(2\pi)^{-\frac{3}{2}}
\int_{\T^3} v(p-p(k)-q)f(q)dq
\no \\
&=[\cE_k(p)-\cE_{\text{min}}(k)
-{\cE}_0(p-p(k))+\cE_{\text{min}}(0)]f(p-p(k)).
\no
\end{align}

Using \eqref{nol} one arrives at  the  representation
\begin{equation}\label{limit}
\Gamma(k) = -\int\limits_{{\bbT}^3}\bigg (\cE_0
(p-p(k))-\cE_{\text{min}}(0)- \cE_{{k}}
(p)+ \cE_{\min}(k)\bigg ) |f (p-p(k))|^2 d p,
\end{equation}
\begin{equation*}
\quad k\in \bbT^3.
\end{equation*}
To check the basic inequality
\eqref{dno} we proceed as follows.

 Making the change of  variable $p\to-p+2p (k)$ in
\eqref{limit} and using the
 fact that  the functions
$\cE_0 (p)$ and $|f(p)|$  are even, one obtains the
representation
 \begin{equation}\label{limit2}
  \Gamma(k)= -\int\limits_{{\bbT}^3}
\big ( \cE_0 (p-p(k))-\cE_{\text{min}}(0)-
 \cE_k (-p+2 p(k))+ \cE_{\text{min}}(k)\big )
 |f(p-p(k))|^2 dp.
\end{equation}

Making again the change of variable $q\to p-p (k)$ in \eqref{limit}
and \eqref{limit2} and adding  the results obtained we get
$$ \Gamma(k)= - \int\limits_{{\bbT}^3}
\cF(k,p) |f(p)|^2 d p,
$$
where
 \begin{equation*}
  \cF(k,p)=\cE_0(p)-\cE_{\text{min}}(0) +\cE_{\min}(k)-
  \frac{ \cE_k(p+p(k))+
\cE_k(p(k)-p)}{2}.
\end{equation*}
By Lemma \ref{twoner}, for any (fixed)
$k\ne 0$, one concludes that  $\cF(k, p)>0$
 for almost
 every $p\in \bbT^3$, proving the basic
inequality \eqref{dno} and the claim follows.

\end{proof}

Our second non-perturbative
  result provides sufficient  conditions for the discrete
spectrum of the whole family of fiber Hamiltonians $h(k)$ with $k\ne0$ to be non-empty.

\begin{theorem}\label{main} Assume Hypothesis \ref{crat}.
 Assume, in addition,  that  the operator $h(0)$  has
 either a threshold eigenvalue
  or a virtual level. Then, for all $k\in \T^3\setminus \{0\}$
  the  discrete spectrum of the fiber Hamiltonian  $h(k)$ below the bottom
  $\cE_{\min}(k)$
  of its
essential  spectrum is a non-empty set.
\end{theorem}

\begin{proof}
The case where $h(0)$ has eigenvalues below the bottom of its
essential spectrum has been already treated in Theorem
\ref{main0}. Assume, therefore, that the  lower bound $m(0)$
 of $h(0)$ coincides with the bottom of its essential
 spectrum, that is,
\begin{equation}\label{pros}
m(0)=\cE_{\text{min}}(0)=\cE_0(0).
 \end{equation}
If, under this hypothesis,
 $\cE_0(0)$ is a (threshold) eigenvalue, the claim  follows from
 Theorem
\ref{main0}.

 Assume, then, that $h(0)$ has a virtual level at the
 bottom of its essential spectrum. Therefore,
 the equation
\begin{equation*}
G(\cE_0(0))\psi=-\psi, \quad \psi\in C(\T^3),
\end{equation*}
 has a nontrivial solution $\psi\in C(\bbT^3)$.
 As in the proof of Theorem \ref{main0},
  without loss of
 generality one may assume that the function $|\psi(p)|$  is even.
In particular, the equation
\begin{equation}\label{nunu}
 {\cE}_0(p)f(p)+(2\pi)^{-\frac{3}{2}}\int_{\T^3}
 v(p-q)f(q)dq=m(0)f(p),
 \quad \text{ a.e. }  p\in \T^3,
 \end{equation}
 has the $L^1(\T^3)$-solution  (cf. Remark \ref{l1l1})
 \begin{equation*}
  f(p)=\frac{\psi(p)}{\cE_0(p)-\cE_0(0)} 
 \end{equation*}
such that the function $|f(\cdot)|$ is even.

 Given  $k\in \bbT^3$,  introduce  the
 sequence $\{ f_{n,k}\}_{n=1}^\infty$ of
$L^2(\bbT^3)$-functions
 \begin{equation*}
f_{n,k}(p)= \frac{\psi(p-p(k))} {{\cE _0} (p-p(k))-\cE_0(0)+ \frac{1}{n}}.
\end{equation*}
By
the dominated convergence Theorem
the sequence $f_{n,k}$ converges in the space $L^1(\T^3)$ as $n\to \infty$
to the function $f_{k}$
$$
f_k(p)=f(p-p(k)), \quad p\in \T^3,
$$
with  $f(\cdot)$  a summable majorant.

Under Hypothesis \ref{hypo} this means that the sequence of functions
$[h(k)-\cE_{\text{min}}(k)]f_{k,n}$
 converges in $L^\infty(\T^3)$-norm to the bounded function
 $$[\cE_k(p)-\cE_{\text{min}}(k)]f_{k}(p)+
(2\pi)^{-\frac{3}{3}} \int_{\T^3} v(p-q)f_{k}(q)dq
$$
$$=[\cE_k(p)-\cE_{\text{min}}(k) -{\cE}_0(p-p(k))+{\cE}_0(0)]f(p-p(k)),
$$
where we used  \eqref{pros},
\eqref{nunu} and   the representations
\begin{align}
&([h(k)-\cE_{\text{min}}(k)]f_{k,n})(p)
\no \\
&=[\cE_k(p)-\cE_{\text{min}}(k)]f_{k,n}(p)+ (2\pi)^{-\frac{3}{3}}
\int_{\T^3} v(p-q)f_{k,n}(q)dq.
 \no
 \end{align}

In particular, one concludes that
the
limit
\begin{equation*}
\Gamma(k)=\lim_{n\to \infty}
  ([h(k)- \cE _{\min}(k)]f_{n,k},f_{n,k}),
  \quad k\in \bbT^3,
\end{equation*}
exists and is finite and, moreover,
\begin{equation*}
\Gamma(k) = -\int\limits_{{\bbT}^3} \frac{\cE_0 (p-p(k))-\cE_0(0)- \cE_{{k}}
(p)+ \cE_{\min}(k)} {(\cE_0(p-p(k))-\cE_0(0))^2}|\psi (p-p(k))|^2 d p,
  \quad k\in \bbT^3.
\end{equation*}

Next,  exactly as it has  been done  in the proof of Theorem
\ref{main0},  one checks
the inequality
\begin{equation}\label{dno1}
\Gamma(k)<0,\quad k\ne 0.
\end{equation}

 It follows from  \eqref{dno1} that there exists an
 $n_0\in \N $ such that
\begin{equation*}
([h(k)- \cE _{\min}(k)]f_{n_0,k},f_{n_0,k})<0, \quad k\ne 0
\end{equation*}
 proving  the existence of the
discrete spectrum of $h(k)$ below  its
 essential spectrum for $k\ne 0$.
The proof is complete.

\end{proof}
\begin{remark}
If the  operator $h(0)$ has a multiple threshold eigenvalue, then
using the same strategy of proof one concludes that
 the variational estimate
\begin{equation}\label{vara}
([h(k)- \cE _{\min}(k)]f_{k},f_{k})<0, \quad k\ne 0,
\end{equation}
with
\begin{equation}\label{fkfk}
f_{k}(p)=f(p-p(k)),
\end{equation}
holds for any $0\ne f\in\Ker(h(0)- \cE
 _{\min}(0)I) $.
Therefore,  estimate \eqref{vara}
holds for all functions $f_k(\cdot)$ of the form \eqref{fkfk}
from a subspace of dimensionality
$$
d=\dim\Ker(h(0)- \cE
 _{\min}(0)I).
$$
This means that  the number $N$ of
eigenvalues of $h(k)$ with $k\ne 0$ (counting multiplicity) below
the bottom of the essential spectrum admits the estimate
$$N\ge \max\{1,d\}.
$$
\end{remark}

\begin{remark} The width $w(k)$ of the essential spectrum band of the
Hamiltonians $h(k)$,
$$
w(k)=\cE_{\max}(k)-\cE_{\min}(k)
$$
may vanish for some values of the quasi-momentum $k\in \T^3$.
Therefore, the fiber Hamiltonians $h(k)$ may have an infinite
discrete spectrum for some values of the quasi-momentum $k$ even
if  the spectrum of $h(0)$ is essential. For instance, consider
two (identical) particles with the
 one-particle dispersion relations of the form
\begin{equation}\label{drdr}
\varepsilon_1(p)=\varepsilon_2(p)=\sum_{i=1}^3(1-\cos p_i), \quad
p\in \T^3.
\end{equation}
Then if $k_0=(\pi,\pi,\pi)\in\T^3$ we
have the ``strong degeneration" of the two-particle dispersion
relation:
$$
\cE_{k_0}(p)=\varepsilon_1(p)+\varepsilon_2(k_0-p)=6\quad \text{
holds for all  } p=(p_1,p_2, p_3)\in \T^3.
$$
 Therefore, in this case,
the fiber Hamiltonian $h(k_0)$ associated with the system of two
particles with dispersion relations \eqref{drdr} interacting via a
fast decreasing potential $v$ with infinitely many negative
eigenvalues has an infinite discrete spectrum below the bottom of
its essential spectrum. (In this particular example the essential
spectrum of $h(k_0)$ is a  one-point set, namely,
$\spec_{\text{ess}}(h(k_0))=\{6\}$).

  It is also worth
mentioning that  even a partial degeneracy of the two-particle
dispersion relation $\cE_k(\cdot)$ for some values of the
quasi-momentum $k\ne 0$
 may generate a ``rich'' infinite discrete spectrum of the Hamiltonian $h(k)$
  outside the band
$[\cE_{\min}(k),\cE_{\max}(k)]$.
 \end{remark}

\appendix
\section{
 Proof of Proposition \ref{ker}}\label{Appendix I}
Assume without loss of generality  that $\varepsilon (0)=0$.

\noindent
 {\it ``Only If Part."} Let $f\in L^2(\T^3)$ be an eigenfunction of the operator $h(0)$
associated with  a zero eigenvalue, that is,
\begin{equation}\label{osn}
-\varepsilon(p)f(p)=(2\pi)^{-\frac{3}{2}} \int\limits_{{\T}^3} v (p-q)f(q)d
q, \quad \text{a.e. } p\in \T^3.
\end{equation}
The same argument as in the proof of Lemma \ref{comp} shows that the equivalence
class associated with the function $f$ has a
 representative $\tilde f$ such that
the function
\begin{equation}\label{psn}
\psi(p)=\varepsilon(p)\tilde f(p)
\end{equation}
is H\"older continuous, $ \psi\in \cB(\kappa) $. Hence the
representative $\tilde f$ is continuous  away from the origin and
since from Hypothesis \ref{hypo} it follows that $ \liminf_{p\to
0}\varepsilon(p)|p|^{-2}>0, $  the following asymptotic
representation
$$
\tilde f(p)=\frac{\psi(0)}{\varepsilon(p)}+\cO(|p|^{-2+\kappa}), \quad p\to 0,
$$
holds.
Since $\tilde f\in L^2(\T^3)$ and $\kappa>\frac{1}{2}$, the H\"older continuous  function $\psi$ must vanish
at the origin, that is,
$
\psi(0)=0.
$

Comparing  \eqref{osn} and \eqref{psn} one concludes that  $-1$ is an eigenvalue
of  the operator $G(\varepsilon(0))$ on $C(\T^3)$  associated with
the eigenfunction $\psi$ with $\psi(0)=0$.

\noindent
{\it ``If Part."} Assume that
 the operator $G(\varepsilon(0))$  has an eigenfunction $\psi$
  associated with the eigenvalue $\lambda=-1$,
  \begin{equation}\label{lll}
  G(\varepsilon(0))\psi=-\psi
 \end{equation}
  such that $\psi(0)=0$.
Following the strategy of proof of Lemma \ref{comp} one gets that
$
\psi\in \cB(\kappa).
$
Introduce the function
$$
f(p)=\frac{\psi(p)}{\varepsilon(p)}, \quad p\ne 0.
$$
Clearly, an argument as above shows that the following asymptotic representation
$$
f(p)=\cO(|p|^{-2+\kappa}), \quad p\to 0,
$$
holds. Since $\kappa >\frac{1}{2}$, one proves that
 $f\in L^2(\T^3)$ and then   \eqref{lll} means that
the operator $h(0)$ has a nontrivial kernel, completing the proof.

\section{Coexistence of a threshold eigenvalue
 and a virtual level}\label{Appendix II}
 
The main goal of this Appendix is to show by an explicit example
that the {\it Case} IV is not empty.
\begin{example}\label{ex2}
Let $\hat h_{\lambda,
\mu}$, $\, \lambda, \mu \in \R,$ be the discrete Schr\"odinger operator
of the form
$$
\hat h_{\lambda,
\mu}=-\Delta+\hat
v_{\lambda,\mu},
$$ where $\Delta$ is the discrete Laplacian from Example \ref{ex}
and
\begin{equation*}\label{poten}
\hat
v_{\lambda,\mu} (s)=
 \begin{cases}
\mu, & s=0\\
\frac{\lambda}{2}, & |s|=1\\
0, & \text{otherwise}.
 \end{cases}
 \end{equation*}
 \end{example}

The Fourier transform of the interaction can be explicitly
computed
\begin{equation*}
 v(p)=\frac{1}{(2 \pi )^{\frac{3}{2}}} \left( \mu +
\lambda \sum_{i=1}^{3}\cos p_i \right),
 \end{equation*}
and, hence, for the Birman-Schwinger kernel one gets the representation
\begin{equation}\label{kernel}
G(p,q;0) =\frac{1}{(2\pi)^3} \frac{\mu +\lambda \sum_{i=1}^3
\cos(p_i-q_i) }{\varepsilon(q)}, \quad p, q \in \T^3,
\end{equation}
where $\varepsilon(q)$ is given by \eqref{disdis} and we have used the equality
$\varepsilon(0)=0$.

Introduce the notations
\begin{align*}
a=&\frac{1}{(2\pi)^{3}} \int_{\T^3}
\frac{dq}{\varepsilon(q)},\quad \quad\, \,\,\,\,
c=\frac{1}{(2\pi)^{3}} \int_{\T^3} \frac{\cos q_idq}{\varepsilon(q)},
\quad i=1,2,3,\no\\
s=&\frac{1}{(2\pi)^{3}} \int_{\T^3} \frac{\sin^2 q_idq}{\varepsilon(q)},
\quad
b=\frac{1}{(2\pi)^{3}} \int_{\T^3} \frac{\cos^2 q_i dq}{\varepsilon(q)},
\,\,\, i=1,2,3,\no \\
d=&\frac{1}{(2\pi)^{3}} \int_{\T^3} \frac{\cos q_i\cos q_j
dq}{\varepsilon(q)},\quad \quad\quad \quad\quad\quad i,j=1,2,3,\,\, i \neq j.\no
\end{align*}

We remark that since  the function
$\varepsilon(q)=\varepsilon(q_1,q_2,q_3)$ is invariant with
respect to the permutations of  its arguments $q_1,q_2$ and $q_3$,
the integrals $c, s, b, d$  above do not depend on the particular
choice of the indices $i, j$. A simple computation shows that the
following relations
\begin{align}
&a-c=\frac{1}{6}, \label{equ0} \\
& b+2d=3c, \label{equ1}\\
 &a=b+s \quad\mbox{and}\quad
s=\frac{1}{6}-\frac{2}{3}(b-d)\label{equ2}
\end{align}
hold.

\begin{lemma}\label{a>} $
a>\frac{11}{51}$ . In particular,
$
c>0$.
\end{lemma}
\begin{proof} We start with the representation
\begin{align}
\int_{-\pi}^\pi \frac{dq}{A-\cos
q}&=\int_{-\frac{\pi}{4}}^{\frac{\pi}{4}} \frac{dq}{A-\cos q}
+\int_{-\frac{\pi}{4}}^{\frac{\pi}{4}} \frac{dq}{A+\cos q} \label{trick}\\
&+\int_{-\frac{\pi}{4}}^{\frac{\pi}{4}} \frac{dq}{A-\sin
q}+\int_{-\frac{\pi}{4}}^{\frac{\pi}{4}} \frac{dq}{A+\sin q},
\quad |A|>1,\nonumber
\end{align}
which yields
\begin{align}
a=\frac{1}{(2\pi)^{3}} &\int_{\T^3}\frac{dq}{\varepsilon(q)}=
\frac{1}{2(2\pi)^{3}} \int_{\T^3}\frac{dq}{3-\cos q_1-\cos
q_2-\cos q_3}
\label{a>=}\\
=\frac{1}{2(2\pi)^{3}} &\int_{\T^2}dq_1dq_2
\int_{-\frac{\pi}{4}}^{\frac{\pi}{4}} dq_3 f(q_1,q_2,q_3),\no
\end{align}
where
\begin{align*} f(q_1,q_2,q_3)=&\frac{1}{3-\cos q_1-\cos
q_2-\cos q_3} +\frac{1}{3-\cos q_1-\cos q_2+\cos q_3}
\\ +&
 \frac{1}{3-\cos q_1-\cos q_2-\sin q_3}+\frac{1}{3-\cos q_1-
 \cos q_2+\sin q_3}.
\end{align*}
Note that the function $f$ is well defined on
$(-\pi,\pi]^3\setminus \{0\}$.

One easily checks that for fixed $q_1, $ $q_2$, the function
$f(q_1,q_2,q_3)$ as a function of the argument $q_3$, $q_3\in
[-\frac{\pi}{4},\frac{\pi}{4}]$, attains its minimal value at the
end points of the interval $[-\frac{\pi}{4},\frac{\pi}{4}]$ and
hence
\begin{equation}\label{first} f(q_1,q_2,q_3)>\frac{2}{3-\cos q_1-
\cos q_2-\frac{\sqrt{2}}{2}} +\frac{2}{3-\cos q_1- \cos
q_2+\frac{\sqrt{2}}{2}},
\end{equation}
$$
q_3\in (-\frac{\pi}{4},\frac{\pi}{4}).
$$
Combining \eqref{a>=} and \eqref{first} proves the inequality
$$
a>\frac{1}{(2\pi)^{3}} \frac{\pi}{2}\int_{\T^2}dq_1dq_2\bigg (
\frac{1}{3-\frac{\sqrt{2}}{2}-\cos q_1-\cos q_2}
+\frac{1}{3+\frac{\sqrt{2}}{2}-\cos q_1- \cos q_2}\bigg ).
$$
Applying the trick \eqref{trick} two more   times  (first by
getting rid
 of the variable $q_2$  and then of  $q_1$) one
arrives at the estimate
\begin{align}
a&>\frac{1}{(2\pi)^{3}} \bigg (\frac{\pi}{2}\bigg
)^2\int_{\T}dq_1\bigg ( \frac{2}{3-2\frac{\sqrt{2}}{2}-\cos
q_1}+\frac{4}{3-\cos q_1} +\frac{2}{3+2\frac{\sqrt{2}}{2}-\cos
q_1}\bigg )
\no \\
&>\frac{1}{(2\pi)^{3}} \bigg (\frac{\pi}{2}\bigg
)^3 \bigg [\frac{4}{3-3\frac{\sqrt{2}}{2}}+
\frac{12}{3-\frac{\sqrt{2}}{2}}+
\frac{12}{3+\frac{\sqrt{2}}{2}}+\frac{4}{3+3\frac{\sqrt{2}}{2}}\bigg
]= \frac{11}{51},\no
\end{align}
completing the proof.
\end{proof}
\begin{corollary}\label{nonempty}
The set
$$
\Lambda=\bigg \{\frac{1}{s},\frac{1}{b-d}\bigg \}\setminus \bigg
\{\frac{2a}{c}\bigg \}
$$
is nonempty.
\end{corollary}
\begin{proof}
 Assume to the contrary that $\Lambda=\emptyset$, that is,
 \begin{equation}\label{final}
\frac{1}{s}=\frac{1}{b-d}=\frac{2a}{c}.
 \end{equation}
 Solving  \eqref{equ1}, \eqref{equ2} and \eqref{final}
 simultaneously in particular yields
 $$
 a=\frac{5}{24}<\frac{11}{51},
 $$
 which is impossible due to Lemma \ref{a>}.
\end{proof}

\begin{theorem}
Assume that $-\lambda\in \Lambda$ and
$$
\mu=-\frac{1+3\lambda c}{a+\frac{\lambda c}{2}},
$$
then the  Hamiltonian $ h_{\lambda,\mu}$ has
 both a virtual level  and a
  threshold
eigenvalue.
\end{theorem}
\begin{proof}

In accordance with Proposition \ref{ker}
one needs to show that the integral operator $G(0)$ given by
\eqref{kernel} on the Banach space $C(\T^3)$  has two
eigenfunctions, $\psi$ and $\varphi$ associated with an eigenvalue
$-1$:
$$
G(0)\psi=-\psi \quad \text{and}\quad  G(0)\varphi=-\varphi
$$
such that
$$
\psi(0)\ne 0 \quad \text{and} \quad \varphi(0)=0.
$$ The space of all odd (resp. even) functions
$C_o({\T^{3}}) $ (resp. $C_e({\T^{3}}) $)
  is an  invariant subspace for the integral operator $G(0)$.
  The restrictions $G_o$ (respectively $G_e$) of $G(0)$
  on the subspace $C_o({\T^{3}}) $ (respectively $C_o({\T^{3}}) $)
  have
 the kernel functions
\begin{align*}
G_o(p,q)&=\frac{ \lambda}{(2\pi)^{3}} \sum_{i=1}^{3} \frac{\sin
p_i \sin q_i }{\varepsilon(q)},\no \\
G_e(p,q)&=\frac{1}{(2\pi)^{3}} \frac{\mu+\lambda
\sum_{i=1}^{3}\cos p_i \cos q_i }{\varepsilon(q)}.
\end{align*}

The matrix of the restriction $G_o\vert_S$ of $G_o$ onto its
three-dimensional invariant subspaces $S\subset C_o(\T^3)$ spanned
by the functions $\sin p_1$, $\sin p_2$, and $\sin p_3$ in the
basis $e_i=\sin p_i$, $ i=1,2,3,$
 is a diagonal matrix of the form
\begin{equation}\label{godd}
G_o\vert_S=
\begin{pmatrix}
\lambda s& 0 & 0\\
0&\lambda s&0\\
0&0& \lambda s
\end{pmatrix}
\end{equation}
while the matrix of the restriction $G_e\vert_C$ of $G_e$ onto its
four-dimensional invariant subspaces $C\subset C_e(\T^3)$ spanned
by the functions $1$, $\cos p_1$, $\cos p_2$, and $\cos p_3$ in
the basis $f_i=\cos p_i$, $ i=1,2,3,$ $f_4=1$  is given by
\begin{equation}\label{geven}
G_e\vert_C=
\begin{pmatrix}
\lambda b  & \lambda d & \lambda d&\lambda c\\
\lambda d&\lambda b&\lambda d&\lambda c\\
\lambda d&\lambda d& \lambda b&\lambda c\\
\mu c&\mu c&\mu c&\mu a
\end{pmatrix}.
\end{equation}

\noindent From \eqref{geven} it  follows that if $\lambda$, $\mu$ and
$\gamma$ satisfy the relations
\begin{equation}\label{eins}
\lambda(b+2d +c\gamma)=-1,
\end{equation}
\begin{equation}\label{zwai}
 \mu(3c+a\gamma)=-\gamma,
\end{equation}
 then
\begin{equation*}\label{reson}
\begin{pmatrix}
\lambda b  & \lambda d & \lambda d&\lambda c\\
\lambda d&\lambda b&\lambda d&\lambda c\\
\lambda d&\lambda d& \lambda b&\lambda c\\
\mu c&\mu c&\mu c&\mu a
\end{pmatrix}
\begin{pmatrix}
1\\1\\1\\\gamma
\end{pmatrix}=\begin{pmatrix}
1\\1\\1\\\gamma
\end{pmatrix}.
\end{equation*}

Given $\lambda\in \R$, $\lambda\ne -\frac{2a}{c}$,  solving
equations \eqref{eins} and \eqref{zwai} with respect to $\mu $ and
$\gamma$
 yields
\begin{align}
\gamma(\lambda)&=-\frac{1}{\lambda c}-3,
\label{gammaeq} \\
\mu(\lambda)&=-\frac{\gamma(\lambda)}{3c+a\gamma(\lambda)}
=\frac{1+3\lambda c}{3\lambda c^2-3\lambda ac -a}=
-\frac{1+3\lambda c}{a+\frac{\lambda c}{2}},
\label{mueq}
\end{align}
where we  used \eqref{equ0} and \eqref{equ1}. Therefore, if
$\lambda\ne -\frac{2a}{c}$ and $\mu=\mu(\lambda)$ satisfies
\eqref{mueq} the operator $G(0)$ has an eigenfunction $
\psi(p)=\gamma(\lambda)+\sum_{i=1}^3\cos p_i$
 and, moreover,
$$\psi(0)=\bigg (\gamma(\lambda) +\sum_{i=1}^3\cos p_i\bigg )\bigg
|_{p_1=p_2=p_3=0}\ne 0
$$
as it follows from \eqref{gammaeq}. Thus, the  Hamiltonian
$h_{\lambda,\mu(\lambda)}$  has a virtual level.

 Next, from the matrix representation \eqref{godd} for $G_o|_S$
 one gets that if
\begin{equation*}\label{ls}
\lambda s=-1,
\end{equation*}
then for any $\mu \in \R$ the operator $G_o$  has a
three-dimensional eigensubspace spanned by the functions $\sin
p_i, $ $i=1,2,3,$
associated with an eigenvalue $-1$ of multiplicity three. In
particular, $G(0)\varphi=-\varphi$
 with $
\varphi(p)=\sin p_1$ and hence $ \varphi(0)=0.$

Similarly (cf. \eqref{geven}), if
\begin{equation*}\label{lbd}
\lambda(b-d) =-1,
\end{equation*}
then for any $\mu\in \R$ the operator $G_e|_C$ has two linearly
independent eigenfunctions $\cos p_1-\cos p_2$ and $\cos p_1-\cos
p_3$ associated with an eigenvalue $-1$ of multiplicity
two. In particular, $ G\varphi=-\varphi$ with $ \varphi(p)=\cos
p_1-\cos p_2$, and hence $\varphi(0)=0.$

Therefore, if
$
\lambda=-\frac{1}{s}
$
or
$
\lambda=-\frac{1}{b-d}
$, then
for any $\mu\in \R$ the operator  $h_{\lambda,\mu}$ has an
eigenvalue at the bottom of its (absolutely) continuous spectrum.

Taking $ -\lambda\in \Lambda=\bigg
\{\frac{1}{s},\frac{1}{b-d}\bigg \}\setminus \bigg
\{\frac{2a}{c}\bigg \} $ (which is nonempty by Corollary
\ref{nonempty}) and $\mu=\mu(\lambda)=-\frac{1+3\lambda
c}{a+\frac{\lambda c}{2}}$ one proves the coexistence of a
virtual level
and a threshold eigenvalue for the Hamiltonian $h_{\lambda,
\mu(\lambda)}$.

\end{proof}
\begin{remark} We were not able to find out whether the set
$\Lambda$ is a one- or a two-point set and hence we cannot
explicitly compute the multiplicity of the zero-energy eigenvalue
(more information about numerical values of the integrals $a, b,
c, $ and $d$ is needed). However,  if $\Lambda$ contains two
elements, then the   Hamiltonian $\hat
h_{\lambda,-\frac{1+3\lambda c}{a+\frac{\lambda c}{2}}}$ has a
virtual level

 and a threshold eigenvalue of multiplicity two or
three depending on the choice of $-\lambda\in\Lambda$
($\lambda=-\frac{1}{b-d}$ or $\lambda=-\frac{1}{s}$ respectively).

 If $|\Lambda|=1$,  it might happen that the Hamiltonian $\hat
h_{\lambda,-\frac{1+3\lambda c}{a+\frac{\lambda c}{2}}}$,
$-\lambda\in \Lambda$, has a virtual level and a threshold
 eigenvalue of multiplicity two, three or
even five depending on which of the cases

 (i) $\frac{c}{2a}=s\ne b-d$;

 (ii) $\frac{c}{2a}=b-d\ne s$;

 (iii) $\frac{c}{2a}\ne s=b-d$;\\
takes place respectively.
\end{remark}

{\bf Acknowledgments} K.~A.~Makarov thanks F.~Gesztesy and
V.~Kostrykin for useful discussions. He is also  indebted to the
Institute of Applied Mathematics of the University Bonn for its
kind hospitality during his stay in the summer 2003. This work was
also partially supported by the DFG 436 USB 113/4 Project and the
Fundamental Science Foundation of Uzbekistan. S.~N.~Lakaev and
Z.~I.~Muminov gratefully acknowledge the hospitality of the
Institute of Applied Mathematics of the University Bonn.

\end{document}